# Measure of Similarity between Fuzzy Concepts for Optimization of Fuzzy Semantic Nets

**Mohamed Nazih Omri[1]  and  Noureddine Chouigui[2]**

[1]ERPAH-FST & [1,2]DMI-IPEIM,
Route de Kairouan, 5019 Monastir.
Tel. 216 3 500 273, Fax. 216 3 500 512
E-mail: Nazih.Omri@ipeim.rnu.tn

**Abstract:** This paper presents a method to measure the similarity between different fuzzy concepts in order to optimize Semantic networks. The problem approached is the minimization of the time of research and identification of user's Objects and Goals. Indeed, it concerns to determine to each instant the totality of Objects (respectively Goals) among which one can identify rapidly the most satisfactory for the user's Object and Goal. Alone Objects and most similar Goals to Objects and researched Goals of the viewpoint of attribute values will be processed, what will avoid the analysis of all Objects and system Goals far of needs of the user.

## 1. Introduction

The Technical System we use has a fuzzy Semantic Networks as a data base and learns by interpreting an unknown word using the links provided by the context of the query and created between this new word and known words. With the learning of new words in natural language as the interpretation, which was produced in agreement with the user, the system improves its representation scheme at each experiment with a new user and, in addition, takes advantage of previous discussions with users.

The Ideal Expert Net of the system we use is defined as the knowledge that is sufficient to the system and that is described in a semantic network [figure1]. Construction of the Ideal Expert Knowledge starts if given a set of Tasks that are executed using elements of one technical Object through procedures. The first step is the task decomposition as a hierarchy of Goal decomposition into sub-Goals from the level of the Goal of the task to primitive actions. The second step consists in (i) drawing up a list of possible Goals and the procedures to reach these Goals (ii) constructing the Ideal Expert Net as a classical semantic network. But, instead of using structural properties of systems interface Objects; Goals reachable with those Objects are used as properties. The ideal user's description uses valid procedures that have to be applied to the elements of the Object in order to successfully complete the task. Classes of Objects and relations between classes of Objects merge from routines for classification and routines for classes organization[22].

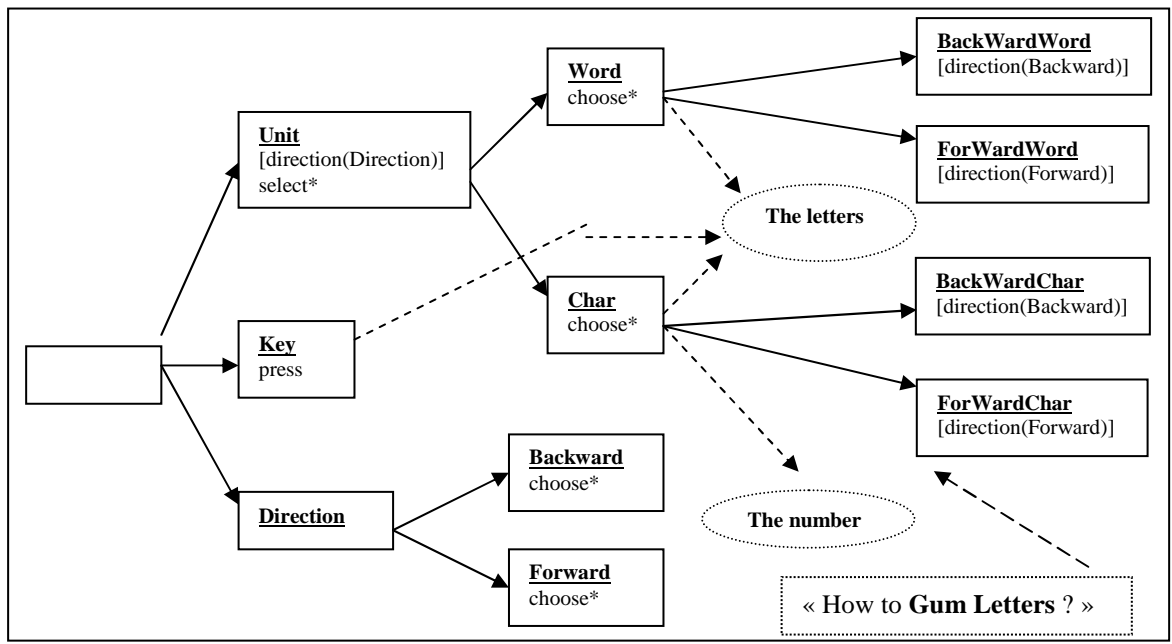

**Figure 1:** The Semantic Network of Novice Users.

## 2. Measure of similarity between fuzzy concepts

The simplification of process, during the automation of the exploitation of knowledge, necessitates an utilization of classify Objects to them limit well defined, attributes admitting precise values and hierarchical bonds non valued, connecting the different Objects of a basis of knowledge, allowing a certain access and without ambiguity to classify them and super - classify an Object given.

Most classic representations in artificial intelligence, by semantic systems and by the logic for example, admit only identifications in all or nothing. The placement in obviousness of the use of resemblance between Objects and the fact to make intervene similarities between Objects met in others situations to solve the problem, are methods taken in account by current researches in artificial intelligence.

## 3. Fuzzy similarity relationship

Let U a non void universe and A, B, C three fuzzy subsets in U.

**Definition 1:** (Zang and al., 91) An application Sim of U x U in [0, 1] is a fuzzy similarity Relationship If, and only If, next properties are verified :

1. Sim (A, A) = 1;
2. Sim (A, B) = Sim (B, A) > 0;
3. If $A(u) \leq B(u) \leq C(u)$ or $A(u) \geq B(u) \geq C(u)$ $\forall$ u $\in$ U, then
   Sim(A, C) $\leq$ Sim(B, C) $\forall$ A, B and C in the universe U.

An example of similarity relationship is given by:
and possesses the next property: Sim (A, B) = 0 If and only if A and B are disjoints.

The degree of similarity has obviously to

$$Sim(A,B) = \frac{\max_{u \in U}(A \cap B)}{\max_{u \in U}(A \cup B)}$$

be calculated between two Objects with similarly nature, generics or individual.

### 3.1. Similarity between two linguistic variables

The degree of similarity between two fuzziness linguistic variable is obtained from values of the degrees associated to it.

Let Y be the universe linguistic values, T and S are two linguistics variables defined on Y such as T = {($t_1$, $d_1$),. ..,($t_n$, $d_n$)} and S = {($s_1$, $d'_1$),. ..,($s_m$, $d'_m$)}. Where $t_i$ and $s_j$, i $\in$ [1, n], j $\in$ [1, m] are the different linguistic variable's values of T and S respectively. $d_i$ and $d'_j$ are the associated degrees to $t_i$ and $s_j$.

**Definition 2:** Let $f^*_{T \cap S}$ and $f^*_{T \cup S}$ are respectively the consequent membership function of the intersection and the union of membership functions and $f^*_{T \cap S}$ associated to T and S. We define the degree of similarity $Sim*$ between T and S by:

This definition applies to the case of possible values and to the case of necessary values. The consequent similarity degree is defined as follows:

$$Sim^*(T,S) = \frac{\max_y f^*_{T \cap S}(y)}{\max_y f^*_{T \cup S}(y)}$$

**Definition 3:** Let $Sim^N$ (T, S) and $Sim^P$ (T, S) respectively necessary and possible similarity degrees between T and S. The consequent similarity degree Sim (T, S) is given by the following definition 4 :

**Definition 4 :** Let T and S are two fuzzy subsets of the universe Y. T and S are perfectly similar if and only if :

$$Sim(T, S) = 1.$$

**Example 1 :** In this example one considers two different descriptions of a same Goal, but belonging to two different Objects 'The-lettres' and 'the-substantif'.

$Erase_{/the\text{-}lettres}$ : { (CutwithMenu, 0.8) (ErasewithKey, 1) (Select, 0.5)}

$Erase_{/the\text{-}substantif}$ : { (ErasewithMenu, 1) (ErasewithKey, 0.7) (Select, 0.5)}

The degree of similarity between the two Goals 'Erase' of the Object 'the-lettres', noted EraLet, and 'Erase' fastened to the Object 'the-substantif', noted EraSub, is calculated as follows:

$$Sim(Erase_{Let}, Erase_{Sub}) = \frac{\max_x f_{Erase_{Let} \cap Era_{Sub}}(x)}{\max_x f_{Erase_{Let} \cup Erase_{Sub}}(x)} =$$

$$\frac{\max(0.8, 0.7, 0)}{\max(1,1,0)} = 0.8$$

### 3.2. Similarity between two fuzzy attributes

The degree of similarity between two attributes will be built in function of similarity degrees between the different values of these attributes and therefore in function of degrees calculated from necessary value zones and possible value degrees associated.

**Definition 5 :** Let $T_i$, $i \in [1, k]$ the set of linguistic values of an attribute A of an Object $O_1$, $S_i$, $i \in <1, k>$, the set of linguistic values of an attribute B of an Object $O_2$. $T_i$ and $S_i$ are defined on the same universe, we define the degree of similarity between A and B by:

$$Sim(A, B) = \min_{1 \leq i \leq k} Sim(T_i, S_i).$$

The degree of similarity between these two Objects is calculated in function of similarity degrees between constitutive attributes of these two Objects.

### 3.3. Similarity between two fuzzy Objects

We come to define the degree of similarity between two fuzzy attributes A and B of two Object $O_1$ and $O_2$. This allows us to obtain the degree of similarity between $O_1$ and $O_2$. It is given by :

**Definition 6 :** Let $O_1$ and $O_2$ two fuzzy Objects respective description $O_1 = \{A_1, ., A_i\}$ and $O_2 = \{B_1, ..., B_n\}$ such that $A_1, B_1, ..., A_i, B_n$ are respectively define on the same universe Y. We define the degree of similarity between $O_1$ and $O_2$ by :

$$Sim(O_1, O_2) = \min_{1 \leq k \leq n} Sim(A_k, B_k).$$

**Example 2 :** We consider the Object *'the-substantive'* and the Object *'the-signs'* whose descriptions by report to the alone possible values are given by the two attributes *'Objects'* indicating the relationship that maintains this Object with the set of Objects and the attributes *'goals'* indicating the set of Goals that we can reach on this Object. These descriptions are following :

**The substantive:**

    Objects: {(The - sign, 1) (the - signs, 0.7) (the - letters, 0.7) (Word, 0.5)}
    Goals: [(to Delete: {( ErasewithMenu, 1) (ErasewithKey, 0.8))
    (to Cut: {( ErasewithKey, 0.7) (ErasewithMenu, 1))]

**The signs:**

Objects: {(The - signs, 1) (The - noun, 0.7) (The - letters, 0.6)]
Goals: [(To delete: {(ErasewithMenu, 1) (ErasewithKey, 0.9))
 (to Cut: {(CutwithMenu, 0.8) (ErasewithKey, 1))]

To calculate the degree of similarity between *'the substantive'* and *'the signs'*, we calculate the degrees of similarity between the different attributes of the two Objects.

$$Sim(LeSubstantif, LesSignes)_{/Objets} =$$

$$\frac{\max(0.7, 0.7, 0.6)}{\max(1,1,0.7)} = 0.7$$

    And

$$Sim(LeSubstantif, LesSignes)_{/Buts} =$$

$$\min\left(\frac{\max(1, 0.8)}{\max(1, 0.9)} + \frac{\max(1, 0.7)}{\max(1,1)}\right) = 1.$$

    Then

$$Sim(LeSubstantif, LesSignes) =$$
$$\min(0.7, 1) = 0.7.$$

In conclusion, one tells that the two Object users are similar with an equal similarity degree to 0,7.

### 3.4. Similarity between two fuzzy instances

Between two fuzzy instances I and I' the degree of similarity calculates the same manner that the degree of similarity between two Objects, in function of similarity degrees between attributes defining these instances.

**Definition 7 :** Are I and I' two blurred authorities respective descriptions I = {$a_1$,..., $a_i$} and I' = {$b_1$,.. .., $b_n$} such that $a_1$, $b_1$,. .., $a_i$, $b_n$ respectively define on the same universe Y. We define a degree of similarity Sim (I, I') between I and I' by:

$$Sim(I, I') = \min_{1 \leq k \leq n} Sim(a_k, b_k).$$

## 4. Construction of classify similar Objects

The central Objective of this paragraph is to regroup in an even classifies objects in order that the value of similarity $Sim(O_i, O_j)$ is at least equal to α. This value has been the chosen threshold in [0, 1]. A class $C^α$ thus constructed will be called class of similar Objects of level α.

The principal Objective of this roundup is to lighten the process of request analysis users presenting a unknown Object. This analysis will be made no longer as compared to the totality of Objects of the basis of knowledge but rather as compared to restrains groups of similar Objects.

### 4.1. Partition classify similar Objects

We have chosen to regroup the totality of Objects of the basis of knowledge some four classify different similarity levels because in reality, it appears that, for more of five levels (0, 0.25, 0.50, 0.75, 1), we have a natural discrimination problem between the different levels and to take less five levels is insufficient to allow a good discrimination. Our Objective is to have an idea more net of the interpretation of a user request as compared to these of the fuzzy knowledge basis.

| Level of Similarity | Corresponding Interval |
|---|---|
| Level 1 | [0, 0.25] |
| Level 2 | [0.25, 0.5] |
| Level 3 | [0.5, 0.75] |
| Level 4 | [0.75, 1] |

Tab. 1 - Different levels of similarity

Let $α_i$, i ∈ [1,4], an interval of level similarity chosen in [0, 1], we regroups in an same classifies Objects in order that the relationship of similarity *Sim* has a value understood between the minimal value $α_{im}$ and the maximal value $α_{iM}$ of this interval $α_i$. The class thus constructed is called class of similarity of level $α_i$.

Let $C_i$, i ∈ [1, 4], the set of classify them similarity of levels $α_i$, one built to each classifies level $α_i$ an Object of reference, noted $O_{ri}$ with respect to the class Here as compared to these Objects of reference, one decides in what class an Object O has to be put. Each Object of reference is built according to the interval of values corresponding to the level similarity. If one considers an Object of reference relative $O_{ri}$ to a class $C_i$, the degrees of possibility that will be associated to values of attributes of $O_{ri}$ will be taken in the interval [$α_{im}$, $α_{iM}$].

If one considers, as an example, the Object relative $O_{ri}$ to the class $C_i$, the degrees of possibility that will be associated to values of these attributes will be in [0, 0.25]. This manner, we are sure and some that Objects of reference thus constructed are different some others and depend directly on the interval associated to the class to the what they make part. One regroups thus in an even classifies Here all Objects whose value of similarity *Sim* with the Object of reference is understood between $α_{im}$ and $α_{iM}$. This condition is formulated by:

$$\forall i \in [1,4], \forall O_{ri}, O_j \in C_i, α_{im} \leq Sim(O_{ri}, O_j) \leq α_{iM}.$$

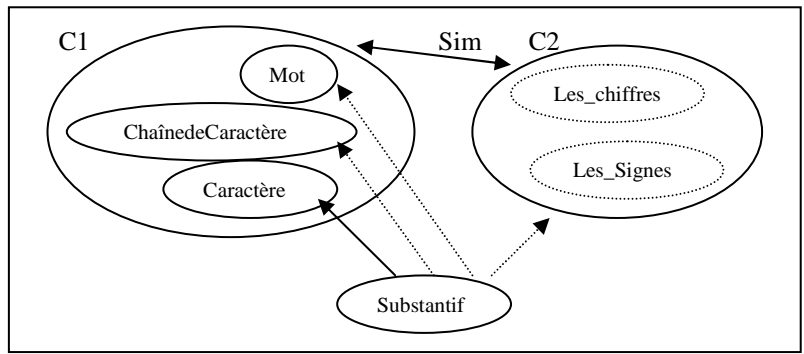

**Figure2:** Partition classify some different levels.

### 4.2. Functioning in Fuzzy Semantic Net

A present unknown Object in the user request has to be analyzed, as compared to all Objects of the Fuzzy Data base, If necessary, so as to identify it. This solution does not appear efficient and rapid in the case where the number of Objects would be very important. This fathers a huge calculation time that will not be necessary, to realize the operation of analysis. Where the partition of this totality in subsets more restrains allowing to minimize the time of research and identification of unknown Objects.

Let a universe X, such that: X = {Character, Word, Chain of Character, Paragraph, Text} whose elements are taken in the framework of a word processing. Is again $C^\alpha$ = {Character, Word, Chain of Character} and $C^\beta$ = {Paragraph, Text} two blurred partitions characterized by classifying them $C^\alpha$ and $C^\beta$ similar Objects of level α and β respectively.

If in the course of the analysis of the user request 'How to Cut a Noun?', the unknown Object 'Substantive' has been identified to the Object system 'Character' belonging to That and that the user is not satisfied, the system presents the rest of Objects belonging to the even classifies it is to tell 'word' and 'Chain of character'. This supposes that Objects similarly class have more chance to be accepted.

If after depletion of all possibilities taken in $C^\alpha$, the user is not always happy, the system has to proceed to the processing of Objects belonging to the class whose level β is the closest to α.

### 4.3. Similarity between two fuzzy Goals

In the case of Goals, the problem is no longer the even because these last present two types of different structures. One distinguishes system's Goals and user's Goals. obviously, the similarity degrees has been made between two similarly typical Goals.

**Case of two system's fuzzy Goals :** Let T and S two linguistics variable defined on the same universe Y. $f_T$ and $f_S$ their respective membership function.

**Definition 8 :** Let $f^*_{T\cap S}$ and $f^*_{T\cap S}$ respectively the consequent membership functions of the intersection and the union of $f_T^*$ and $f_S^*$, we defines the degree of similarity Sim * between T and S by :

$$Sim^*(T,S) = \frac{\max_y f^*_{T\cap S}}{\max_y f^*_{T\cup S}}.$$

This definition applies to the case of possible values and to the case of necessary values where :

$$Sim^*(B_1, B_2) = min_i\ Sim^*(T_i, S_i).$$

The consequent similarity degree is defined as follows:

$$Sim(B_1, B_2) = \frac{Sim^N(B_1, B_2) + Sim^P(B_1, B_2)}{2}.$$

**Case of two user's fuzzy Goals :** As indicated previously, the structure of a user's Goal does not present necessary type values. Values that constitute a user attribute are possible type. The degree of similarity between two user's fuzzy Goals is calculates therefore in the same manner that the degree of similarity between two system's fuzzy Goals, but by canceling the necessary quantity and to take in account only the possible quantity.

$$Sim(B_1, B_2) = min_i\ Sim(T_i, S_i).$$

With $\quad Sim(T_i, S_i) = \dfrac{\max_y f_{T_i \cap S_i}(y)}{\max_y f_{T_i \cup S_i}(y)}.$

### 4.4. Construction of classify similar Goals

The same step applied in the case of classify them similar Objects is applied in the case of Goals for the construction of classify similar blurred Goals.

### 5. Conclusion

The approach presented in this paper, that consists of a construction of sets of similar Goals and Objects in order to optimization fuzzy semantic Networks, does not represent a general methodology to diagnosis the goal query's novice users, allows identifying the unknown novice user's query. This can serve as basis for our research so as to elaborate a general methodology to diagnosis the purpose Goal of the subject, applicable to a large diversity of Objects. The development of this method would have to allow a best approximation of the category of the purpose aimed by the user, and best approaches the diagnosis. This makes only increase performances of the system in the course of the identification of user requests.